\documentclass{article}

\usepackage[final]{changes}
\usepackage{amsmath}
\usepackage{xspace}
\usepackage{float}
\usepackage{authblk}
\usepackage[numbers,super]{natbib}
\usepackage{graphicx}
\usepackage{lineno}
\usepackage{array}
\usepackage[normalem]{ulem}
\usepackage[margin=1in]{geometry}

\providecommand{\eprint}{}

\providecommand{\adsurl}[1]{\href{#1}{ADS}}
\def\aap{{Astron. Astrophys.}}		
\def\apj{{Astrophys. J.}}			
\def\aj{{Astron. J.}}
\def\apjl{{Astrophys. J. L.}}		
\def\apjs{{Astrophys. J. S.}}		
\def\pasp{{Publ. Astron. Soc. Pacif.}}		

\def\mnras{{Mon. Not. R. Astron. Soc.}}
\def\nat{{Nature}}

\def\aaps{{Astron. Astrophys. Suppl. Ser.}}

\newcommand{\microns}{$\mu$m}

	\title{UV Absorption by Silicate Cloud Precursors in Ultra-hot Jupiter WASP-178b}

	\author[1,2]{Joshua D. Lothringer*\thanks{These authors contributed equally to this work. *e-mail: {jlothringer@uvu.edu}; \mbox{dsing@jhu.edu}}} 
	\affil[1]{Department of Physics, Utah Valley University, Orem, UT 84058, USA}
	\affil[2]{Department of Physics and Astronomy, Johns Hopkins University, Baltimore, MD, USA}

	\author[2,3]{David K. Sing*}
	\affil[3]{Department of Earth \& Planetary Sciences, Johns Hopkins University, Baltimore, MD, USA}
	
	\author[3]{Zafar Rustamkulov}
	\author[4]{Hannah R. Wakeford}
	\affil[4]{School of Physics, University of Bristol, HH Wills Physics Laboratory, Tyndall Avenue, Bristol BS8 1TL, UK}
	
    \author[5,3]{Kevin B. Stevenson}
    \affil[5]{Johns Hopkins University Applied Physics Laboratory, Laurel, MD 20723, USA}
    
    \author[6]{Nikolay Nikolov}
    \affil[6]{Space Telescope Science Institute, 3700 San Martin Drive, Baltimore, MD 21218, USA}
    
    \author[7]{Panayotis Lavvas}
    \affil[7]{Groupe de Spectrométrie Moléculaire et Atmosphérique, Université de Reims, Champagne-Ardenne, CNRS UMR F-7331, France}

	\author[8]{Jessica J. Spake}
	\affil[8]{Department of Astronomy, California Institute of Technology, Pasadena CA 91125, USA}
	
	\author[9]{Autumn T. Winch}
	\affil[9]{Department of Physics, Bryn Mawr College, Bryn Mawr, PA, USA}

\begin{document}
\renewcommand{\figurename}{\textbf{Fig.}}
\setcounter{figure}{0}   

\renewcommand{\tablename}{\textbf{Table}}
\setcounter{table}{0}  

\maketitle
\textbf{Aerosols have been found to be nearly ubiquitous in substellar atmospheres\citep{cushing:2006,saumon:2008,burningham:2021}. The precise temperature at which these aerosols begin to form in exoplanets has yet to be observationally constrained. Theoretical models and observations of muted spectral features suggest that silicate clouds play an important role in exoplanets between at least 950 and 2,100 K \citep{gao:2020}. However, some giant planets are thought to be hot enough to avoid condensation altogether\citep{lothringer:2018,kitzmann:2018}. Here, we present the near-UV transmission spectrum of an ultra-hot Jupiter, WASP-178b ($\sim$2,450~K), that exhibits significant NUV absorption. This short-wavelength absorption is among the largest spectral features ever observed in an exoplanet in terms of atmospheric scale heights. Bayesian retrievals indicate the presence of gaseous refractory species containing silicon and magnesium, which are the precursors to condensate clouds at lower temperatures. SiO in particular has not been detected in exoplanets \added{before}, but the presence of SiO in WASP-178b is consistent with theoretical expectation as the dominant Si-bearing species at high temperatures. These observations allow us to re-interpret previous observations of HAT-P-41b and WASP-121b that did not consider SiO to suggest that silicate cloud formation begins on exoplanets with equilibrium temperatures between 1,950 and 2,450~K.}

We observed one transit of the ultra-hot Jupiter WASP-178b/KELT-26b\citep{hellier:2019,rodriguezmartinez:2020} (WASP-178b hereafter) with HST/WFC3/UVIS using the G280 grism (0.2-0.8~\microns{}, R$\sim$70) on September 5th, 2020 as part of Program 16086 (PI: Lothringer). WASP-178b is unique among known exoplanets for its especially hot host star: at A1 IV-V and T$_{\textrm{eff}}$=9360 K, WASP-178 is second-only to KELT-9 as the hottest planet-hosting star. Time-series spectra were obtained over 7.5 hours centered around the transit event, and were used to extract the transmission spectra which probes the middle and upper atmosphere around the day-night terminator.  Further details regarding the data reduction and observational setup are given in the Methods. 

\begin{figure*}[h]
	\centering
	\includegraphics[width=6.5in]{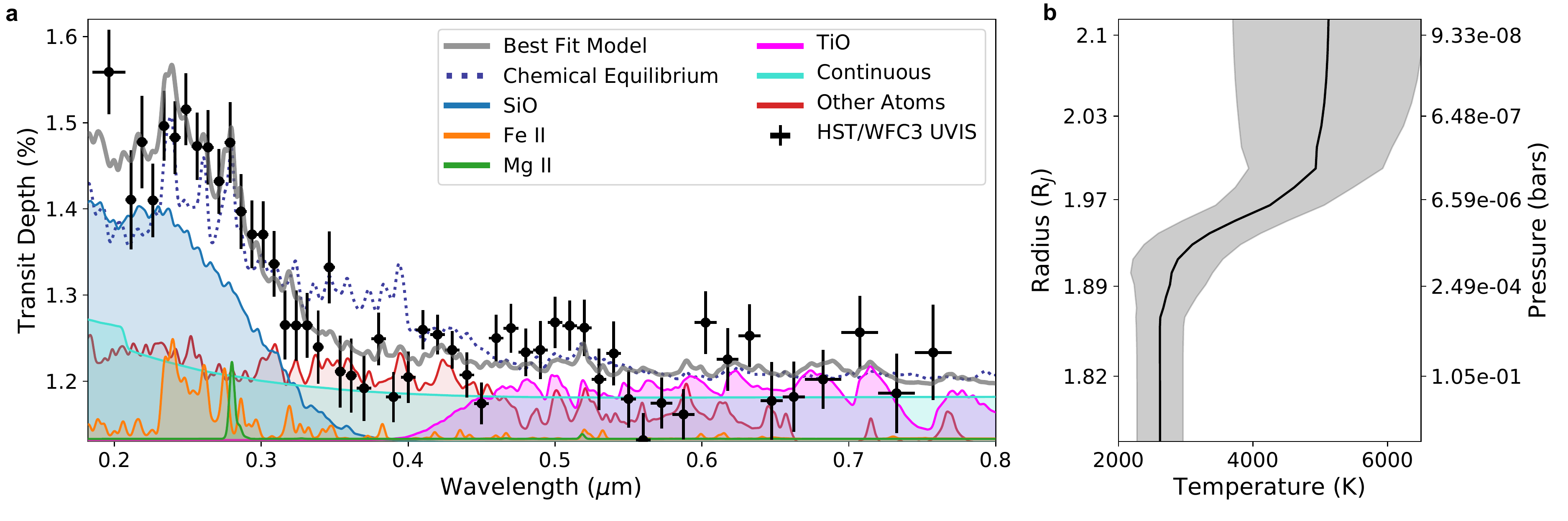} 
	\caption{\textbf{WASP-178b NUV-Optical Transmission Spectrum.} \textbf{a.} WFC3/UVIS G280 transmission spectrum of WASP-178b (with 1-$\sigma$ uncertainties) compared to \replaced{the contribution of various important opacity sources in the retrieved best fit spectrum}{the best fit retrieved model, as well as the transit depths of an atmosphere made of individual opacity sources}. SiO dominates the contribution at short wavelengths. \textbf{b.} The 1-$\sigma$ constraint on the pressure-temperature profile (shaded region) with the median retrieved profile (solid line) on the same radius scale as the left plot, indicating the minimum pressures probed by our observations are less than a microbar. \label{fig:modelcompare}}
\end{figure*}

The resulting spectrum is shown in Fig.~\ref{fig:modelcompare}. A steep rise in transit depth is seen toward short wavelengths, beginning at about 0.35~\microns{}. The difference in NUV transit depth (0.2-0.28~\microns{}) compared the optical transit depth (0.35-0.8~\microns{}) is 2,500$\pm$138 ppm, an 18.0$\sigma$ significance. In terms of the equilibrium atmospheric scale height (calculated at the equilibrium temperature and assuming an H$_2$-dominated atmosphere), the NUV transit depths rise nearly 20 scale heights above the optical continuum, making this one of the largest known spectral features yet seen in an exoplanet atmosphere. \added{A lack of transit asymmetry indicates that this absorption is present on each limb (see Extended Data Fig.~\ref{fig:limb_asym} and Methods).}\added{ We are able to rule out a scattering slope from, e.g., photochemical hazes, as well as stellar inhomogeneities as causes for this feature (see Methods). We are thus confident that the rise in the transit depths at NUV wavelengths is due to absorption by gaseous species in the atmosphere of WASP-178b.}

We ran a series of retrievals to obtain constraints on the atmospheric properties from the observations \added{(see Table~\ref{tab:retrievals})}. The retrievals included free parameters for the abundance of major NUV and optical absorbers, including SiO, TiO, VO, Fe I, Fe II, Mg I, and Mg II, plus a general [Fe/H] parameter for the abundance of all other atmospheric species. We also used a five-parameter temperature structure parameterization (described in Methods). Fig.~\ref{fig:modelcompare} shows the retrieved best fit spectrum\added{ from our fiducial retrieval with all opacity sources present} and the contribution from \added{these }various opacity sources. Also plotted are the constraints on the temperature structure. The maximum transit depths of about 1.5\% correspond to a radius of about 2~$R_\textrm{Jupiter}$. This corresponds to a pressure of about a microbar, similar to the strong lines of Na and K at high spectral resolution. 

In all the retrievals we tested, two scenarios were able to fit the data: 1) an atmosphere with an approximately solar abundance of SiO or 2) an atmosphere with a super-solar abundance of Mg I and Fe II\added{ but with no SiO}. SiO absorbs throughout the 0.2-0.35~\microns{} range, enabling a good fit to the data. On the other hand, bound-free opacity from Mg I absorbs shortward of 0.255~\microns{} \citep{matsushima:1968,fontenla:2015}, while Fe II absorbs between 0.24 and 0.3~\microns{}. Thus the combined absorption from Mg I and Fe II is also able to provide an adequate fit to the observations, albeit at super-solar abundances \added{(see Extended Data Fig.~\ref{fig:modelcompare_nosio})}. Taken in tandem, our analysis indicates that there is strong evidence ($\Delta{BIC}=8.39$) that SiO or Mg must be present in the atmosphere of WASP-178b to explain our observations. Since both Mg and SiO are the major constituents of silicate condensates like enstatite (MgSiO$_3$) and forsterite (Mg$_2$SiO$_4$), we can say with equally high confidence that silicates have not \replaced{rained out}{condensed} at the terminator of WASP-178b. \added{Owing to the lack of transit asymmetry mentioned above, this result holds for both the evening and morning terminator.}

This result is in line with theoretical expectations. Previous studies have pointed out that above 2,000~K, SiO is expected to be a major absorber shortward of 0.4~\microns{} at its chemical equilibrium abundance \citep{sharp:2007,lothringer:2020b} and so we would expect to see it in the transit spectrum \textit{a priori} (also see Extended Data Fig.~\ref{fig:si_pp}). Similarly, we expect many neutral and ionized atomic species to be present, as indicated by chemical equilibrium calculations and high-resolution observations \citep[e.g.,][]{hoeijmakers:2020b}. While SiO and Mg I + Fe II independently provide good fits to the data, \deleted{we consider the most likely scenario to be a combination of these opacity sources}\added{our prior expectation based on chemical equilibrium is that each of these opacity sources are likely present}.

The non-detection of neutral and ionized Fe is somewhat surprising since the species has been detected in planets of similar temperature, like KELT-20b/MASCARA-2b\citep{stangret:2020}, WASP-76b\citep{ehrenreich:2020,kesseli:2021}, WASP-121b \citep{sing:2019,gibson:2020,hoeijmakers:2020b,cabot:2020}. The apparent absence of significant Fe~I absorption in WASP-178b's spectrum could be explained by the high temperatures and high UV flux from the A-type host star ionizing most of the Fe~I. Ground-based high-resolution studies of WASP-178b could provide a comprehensive census of neutral and ionized refractory species, as has been done for other ultra-hot Jupiters \citep{hoeijmakers:2019,hoeijmakers:2020b,merritt:2021}.

To spectrally resolve potentially escaping neutral and ionized Fe and Mg features, we additionally analyzed high resolution NUV transit data of WASP-178b HST data taken with STIS/E230M (see Methods). The STIS data shows no evidence for either Fe II or Mg II (see Extended Data Fig. \ref{fig:E230M}), even though both elements were easily detected in similar data of WASP-121b\cite{sing:2019}. The STIS E230M transmission spectrum is in good agreement with the broadband UVIS spectrum, indicating unresolved escaping Fe II and Mg II lines are not the cause of the NUV absorption feature in the UVIS spectrum, with continuum level absorption by SiO, Mg I, and Fe II the most likely scenario.

\begin{figure*}[h]
	\centering
	\includegraphics[width=6.0in]{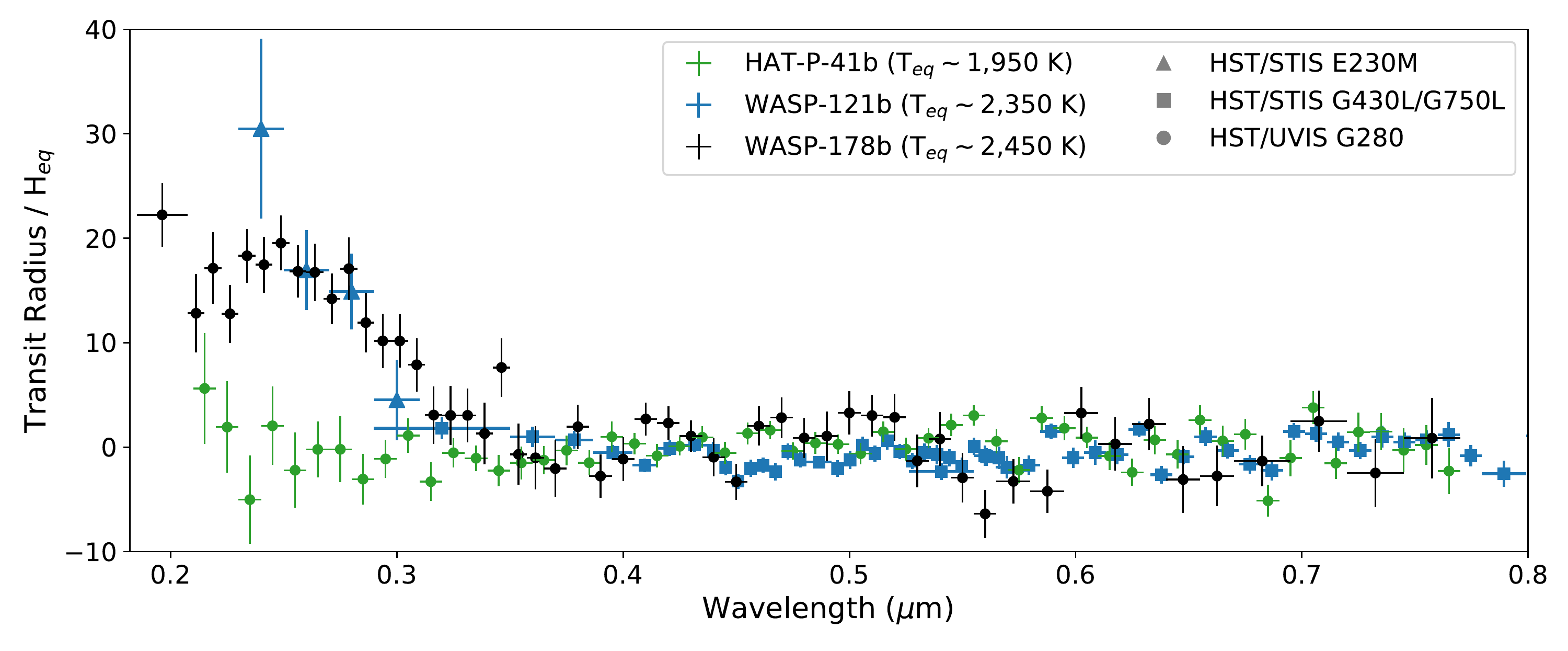} 
	\caption{\textbf{Comparison of NUV-Optical Transmission Spectra.} WFC3/UVIS G280 transmission spectrum of WASP-178b (with 1-$\sigma$ uncertainties) compared to the UV and optical spectra of similar giant planets HAT-P-41b \citep{wakeford:2020AJ} and WASP-121b \citep{evans:2018,sing:2016}, normalized by each planets equilibrium temperature scale height. Significant UV absorption is seen at the shortest wavelengths in WASP-121b and WASP-178b. \label{fig:UHJcompare}}
\end{figure*}

Only a handful of observations exist that are precise enough to measure the continuum of exoplanets shortward of 0.35~\microns{}: HAT-P-41b (T$_{eq}$ = 1950~K) has been observed with a similar setup to our observations of WASP-178b (i.e., HST/WFC3/UVIS/G280)\citep{wakeford:2020AJ}, while WASP-121b (T$_{eq}$ = 2350~K) has been observed at high-resolution with HST/STIS/E230M with 4 binned points between 0.23 and 0.31~\microns{}\citep{sing:2019}.
A clear difference between the spectrum of HAT-P-41b and those of the hotter WASP-121b and WASP-178b is apparent (see Fig.~\ref{fig:UHJcompare}). While WASP-121b's transit spectrum indicates a similar level of absorption at NUV wavelengths to WASP-178b, HAT-P-41b's spectrum shows a definite absence of absorption\added{ on both limbs} at these same wavelengths. This dichotomy suggests that while gaseous refractory species like SiO, Mg, and Fe are abundant in the atmospheres of WASP-121b and WASP-178b, such species have rained out of the gaseous phase in HAT-P-41b. \replaced{Because some of the NUV absorption in WASP-121b is from escaping exospheric metals \citep{sing:2019} and the difference between the morning and evening terminator has not been well-constrained like WASP-178b, we choose to define the equilibrium temperature of WASP-178b as the empirical upper limit for the onset of silicate condensation in hot Jupiters. Therefore, silicate condensation at the terminators must begin between equilibrium temperatures of 1,950~K and 2,450~K.}{We interpret this as \deleted{direct }evidence that iron and silicate species condense in hot Jupiters at equilibrium temperatures between 1,950~K and 2,350~K. While WASP-121b provides the upper bound to this temperature constraint, it is the new, better-resolved, more-precise, and less ambiguous WASP-178b observations and their retrievals that enable this interpretation.} 

\begin{figure*}[h!]
	\centering
	\includegraphics[width=4in]{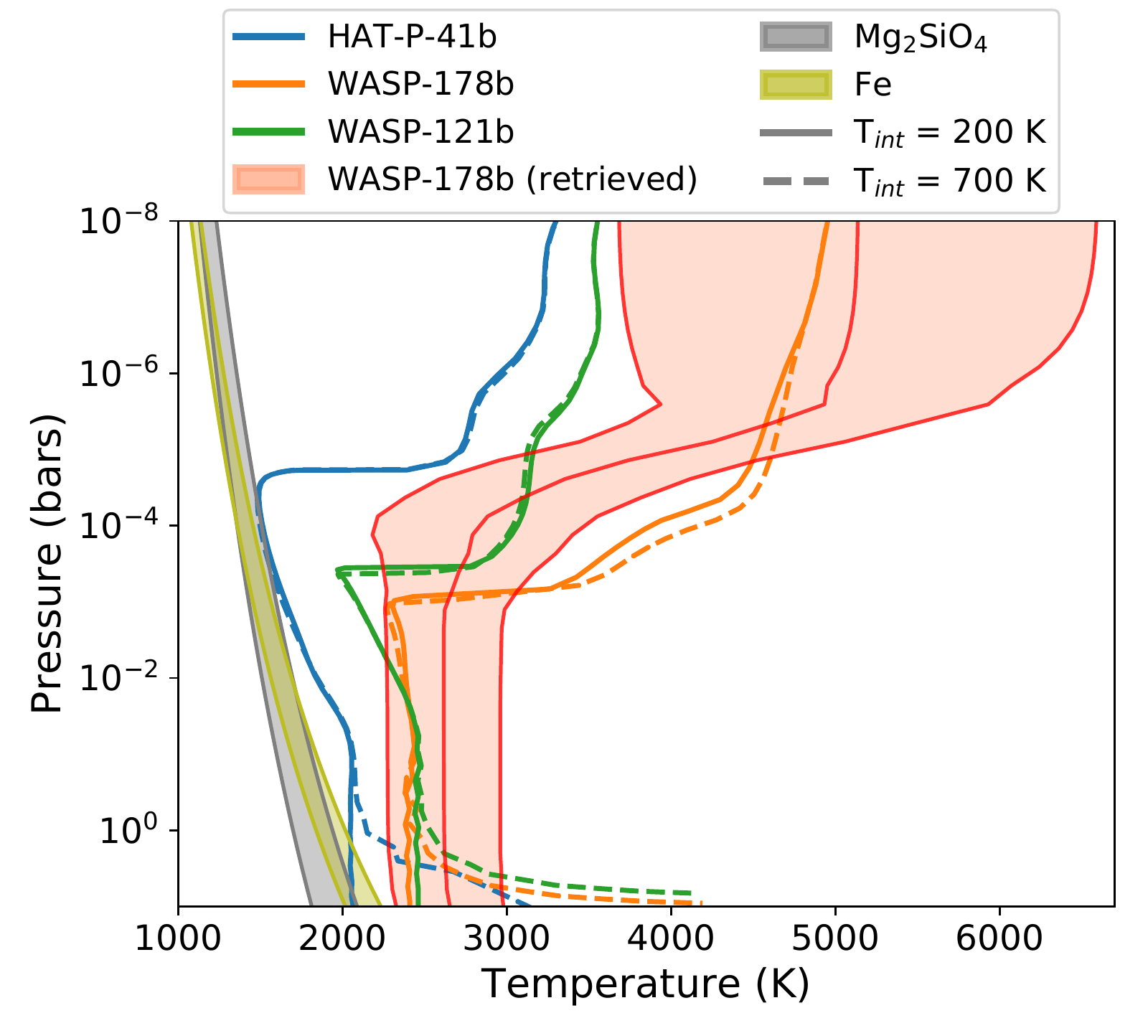} 
	\caption{\textbf{Atmospheric structures and Condensation Curves.} Pressure-temperature profiles of three ultra-hot Jupiters from atmosphere models (see Methods) compared to the condensation curve of Fe and Mg$_2$SiO$_4$\added{ in a 1-10$\times$ solar metallicity atmosphere} \citep{visscher:2010}.\deleted{ The lower temperature limit of the condensation curve is the condensation temperature in a 1$\times$ solar metallicity atmosphere, while the upper temperature limit is the condensation temperature in a 10$\times$ solar metallicity atmosphere.} Where the profile of the planets intersects the condensation curve, as in the case of HAT-P-41b, is where clouds are expected to form in equilibrium. We also show the retrieved median and 1-$\sigma$ confidence interval for the 1D pressure-temperature profile of WASP-178b for comparison, \replaced{demonstrating}{showing} agreement with self-consistent atmosphere model expectation. \label{fig:pp_cond_comapre}}
\end{figure*}

This empirical constraint on the onset of condensation is \replaced{consistent with}{very similar to} theoretical predictions \citep{visscher:2010,gao:2020}. Fig.~\ref{fig:pp_cond_comapre} compares pressure-temperature profiles from theoretical 1D atmosphere models of HAT-P-41b, WASP-121b, and WASP-178b to condensation curves of silicate and iron species. In equilibrium, silicates and iron will condense between about 1500 and 2000~K between 1 mbar and 10 bar for atmospheric metallicities between 1x and 10x solar. \added{Throughout the atmosphere, HAT-P-41b is much closer to the silicate condensation curve than WASP-121b and WASP-178b and will almost certainly cross it on the nightside. If WASP-121b and WASP-178b do reach temperatures cool enough to condense silicates on the nightside, it also appears they are both able to avoid rainout on either limb through rapid evaporation, vertical lofting, insufficiently rapid nucleation and condensation, or some combination of these and other hydrodynamic and microphysical processes\added{ \citep{parmentier:2021,roman:2021,helling:2021}}. At depth, where the temperatures in WASP-121b and WASP-178b are the closest to the condensation curves, the higher \deleted{equilibrium }internal temperature in hot and ultra-hot Jupiters may also help such planets to avoid condensation \citep{thorngren:2019}.}\deleted{At both high and low pressures, the atmosphere of HAT-P-41b reaches temperatures where silicates and iron will condense. WASP-121b and WASP-178b, on the other hand, remain just hot enough to stay above the condensation temperatures at pressures below 10 bar. We also found that the retrieved temperature profile of WASP-178b from the HST/WFC3/UVIS observations matches the 1D theoretical profile well, also remaining above the condensation curves of silicates and iron. At depth, where the temperatures in WASP-121b and WASP-178b are the closest to the condensation curves, the higher \deleted{equilibrium }internal temperature in hot and ultra-hot Jupiters may also help such planets to avoid condensation \citep{thorngren:2019}.}

\added{As noted above, few NUV transit spectra exist for hot Jupiters. Future low- and high-resolution observations, combined with multi-dimensional theoretical modeling \citep{parmentier:2021,roman:2021,helling:2021} and lab studies of aerosols \citep{horst:2018,fleury:2019} in hot and ultra-hot Jupiter, could provide more detailed constraints on the beginning of cloud formation in these atmospheres, while taking into account the myriad processes that promote or inhibit cloud formation, such as night-side cold-trapping, rainout, and vertical mixing and other potentially confounding variables like surface gravity and host star type. We estimate about two dozen planets Jovian exoplanets can be characterized with HST/WFC3-UVIS/G280 with four or less transits \citep{kempton:2018,mullally:2019}.\added{ If combined with STIS/E230M observations to disentangle the effects of atmospheric escape like we have done here, these planets can reveal the conditions for and sequence of condensation in exoplanet atmospheres.} Observing, modeling, and retrieval of brown dwarfs at similar effective temperatures will also shed light on these questions \citep{luna:2021,burningham:2021}.}

\begin{table*}[h!] 
		\centering  
		\caption{Retrieval Results Summary}
		\label{tab:retrievals} 
		\begin{tabular}{ccccc}
			\hline    
			\hline 
			Scenario & $N_\textrm{params}$ & Max. ln($\mathcal{L}$) & $\chi^2_{\nu}$ (K) & $\Delta$BIC \\
			\hline
			Full & 14 & 353.41 & 1.44 & -- \\
			No Mg II & 13 & 352.13 & 1.47 & -1.447 \\
			No Fe II & 13 & 351.13 & 1.51 & 0.56 \\
			 No SiO & 13 & 351.10 & 1.52 & 0.61 \\
			 No SiO or Mg I & 12 & 345.21 & 1.75 & 8.39 \\
			 No SiO or Fe II & 12 & 343.92 & 1.81 & 10.965 \\
			 Full w/ Haze & 16 & 352.15 & 1.58 & -10.53\\
			\hline 
		\end{tabular}

	\end{table*} 

\clearpage

\section*{Methods}
\textbf{HST/WFC3 observations}
One transit was observed of WASP-178b with the HST/WFC3/UVIS instrument using the G280 grism (0.2-0.8~\microns{}). The transit is covered over five HST spacecraft orbits, with the transit approximately centered in the third orbit. Exposure times of 40 seconds were used, along with a 590$\times$2250 pixel detector sub-array which reduced the readout overheads providing 123 total exposures, with 23 to 25 exposures per HST orbit. The spectrograph is slitless, and we centered the subarray such that both the +1 order and -1 order spectra were recorded and could both be fully analysed. Utilizing both orders provides two independent transmission spectra of the same transit event, though the +1 order provides higher signal-to-noise (SNR) given a higher throughput (see Extended Data Fig.~\ref{fig:orders}). Also see ref.~\citep{wakeford:2020AJ} for more information on the instrument mode and analysis.
\\

\textbf{Data Reduction} 
The raw data was processed with the STScI CALWF3 pipeline (v 3.5.1) which applies reduction steps including bias subtraction, dark correction, and flat fielding. The target flux was subsequently extracted starting from the pipeline FLT files.  From each image we used the mode to measure and subtract the background flux. We then removed cosmic rays following a two step process. First, we identified and removed cosmic rays utilizing the time series counts of each pixel. Outlier cosmic rays were flagged and replaced with a 5-$\sigma$ clipping algorithm. We then removed cosmic rays spatially, using a Laplacian edge detection algorithm on each image separately \citep{2001PASP..113.1420V}. We then extracted the 1D spectral flux for each image on both the -1 and +1 orders separately using IRAF APALL with an 8$^{th}$ order Legendre polynomial fit to the spectral trace.  A large range of aperture sizes were extracted, between 10 and 28 pixels, with a 14 pixel aperture found to be optimal in the subsequent light curve fitting stage. The wavelength solution was determined from the spectral trace detector position following ref.~\citep{2017wfc..rept...20P}. 
\\

\textbf{UVIS Light Curve Analysis} 
The general light curve fitting followed the procedures detailed in Sing et al. (2019) \citep[][]{sing:2019} and previously used on WFC3/G280 observations in Wakeford et al. (2020)\citep{wakeford:2020AJ}, which we refer the reader to for subsequent details. The flux measurements over time, $f(t)$, were modeled as a combination of a theoretical transit model\cite{mandel:2002}, $T(t, \mathbf{\theta})$
(which depends upon the transit parameters $\mathbf{\theta}$), the total
baseline flux of the star, $F_0$, and a instrument systematics 
model $S(\mathbf{x})$ giving,
\begin{equation} 
f(t)=T(t, \theta)\times F_0 \times S(\mathbf{x}).
\end{equation} 
As in ref. \cite{sing:2019}, we explored a wide range of models for $S(\mathbf{x})$, exploring detrending variables including a fourth order polynomial in HST orbital phase and linear terms in spectral position as measured from the spectral extraction, wavelength shift measured from cross-correlation of each spectra, and the spacecraft Jitter Detrending vectors which are products of HST's Engineering Data Processing System. For each both the +1 and -1 orders, we used the Akaike information criterion \added{(AICc) with a correction for small sample sizes} to determine the optimal detrending variable parameters to include from the full set without overfitting the data and while minimizing the rednoise. \added{The light curve error bars were derived from the residual scatter of the best fit. In addition, residual systematic noise, $\sigma_r$, was measured along with the white noise, $\sigma_w$, using the binning technique\cite{pont:2006}, with the final fit parameter errors inflated by a factor $\beta$ if rednoise was present\cite{winn:2007} (see Extended Data Table~\ref{Table:data}). Overall, the effect of rednoise was minimal as the noise was comparable to the binned photon noise at typically several hundred ppm, and the errors in only 5 of 124 light curves required an increased scaling by more than 10\%.} 

To model the effects of limb-darkening and center-to-limb differences in the star during transit, we calculated a custom PHOENIX stellar model using the parameters of WASP-178A. For each wavelength bin of interest, we then used the stellar intensity profile to fit for limb-darkening coefficients using the non-linear 4-parameter limb-darkening as described in ref. \cite{2010A&A...510A..21S} which were subsequently used in the light curve transit fit. As a test of the stellar models, we additionally fit for the limb-darkening assuming a linear law between the wavelengths of 0.255 and 0.285$\mu$m. At these NUV wavelengths, the stellar limb-darkening is strong, but also predicted by models to be nearly linear in intensity across the limb, allowing for a direct comparison largely free of complex degeneracies between transit-fit coefficients. We found the transit data fit with a linear coefficient of $u=0.778\pm0.028$, which matches very well to the PHOENIX model prediction of $u=0.7758$. The limb-darkening was subsequently fixed to the PHOENIX model values for all the transit light-curve fits.

We first fit the white light curve, which integrates the entire spectra (see Extended Data Fig.~\ref{fig:wlc}). The planet's orbital system parameters including inclination, $i$, semi-major axis in units of stellar radii, $a/R_{star}$, and time of transit $T_0$ were fit along with the planet-to-star radius ratio, $R_{pl}/R_{star}$, $F_0$, and systematics model using the +1 order.  Given the good phase coverage, these system parameters were generally more accurate over previous literature values finding $i=84.41\pm0.20$ degrees, $a/R_{star}=6.588\pm0.091$, and $T_0=2459097.869279\pm0.00014$ days. We kept the period fixed to the literature value of 3.3448285$\pm$0.0000012~days\citep{hellier:2019}.

To derive the planetary transmission spectrum, we fixed the system parameters to the white-light curve best-fit values and used the optimal systematics model $S(\mathbf{x})$ as determined by the white-light curve fit. While the white-light curve did not require a term to model wavelength shifts in the time-series spectra, this term was found to be needed in the spectral light curve fits with 10 fit parameters overall for each light curve. \added{The residual fit scatter in the time-series spectral bins achieved a level that was on average 1.2$\times$ theoretical photon-noise limit scatter and typically ranged between 1.1$\times$ and 1.4$\times$ (see Extended Data Fig.~\ref{fig:lcbins}).} A variety of spectral bin locations and resolutions were measured, with the adopted spectra chosen to balance the resolution and SNR. In each case, the overall shape of the transmission spectrum was consistent between different resolutions. We measured the transmission spectrum of both the +1 and -1 orders, finding good agreement between both orders (see Extended Data Fig. \ref{fig:orders}). We calculated the weighted-mean value of the spectra to report our final derived spectrum (see Extended Data Table~\ref{Table:data}) as seen in Fig. \ref{fig:modelcompare}. 

We also independently verified the transmission spectrum using the marginalization method described in \citep{wakeford:2020AJ}. This method resulted in a spectrum that was consistent with the spectrum analyzed with Jitter Detrending, both showing large NUV absorption. This method used limb darkening coefficients from the Kurucz stellar model grid \citep{kurucz:1992}, again suggesting our results are robust against the details of limb darkening.
\\

\textbf{STIS E320M Light Curve Analysis} 
To help resolve possible Fe II and Mg II features in the NUV spectrum of WASP-178b, we also analysed a transit observed on Jul 30, 2020 by HST with the STIS E230M instrument. 
As with the UVIS data, these observations were also taken as part of program 16086. The STIS observations were observed with the NUV-MAMA detector using the Echelle E230M STIS grating
and a square 0.2$''\times0.2''$ aperture.
The E230M spectra has a resolving power of R=$\lambda$/(2$\Delta\lambda$)=30,000 and we set
the grating to 2707\AA\ to cover the wavelength ranges from 2280 to 3070 \AA\ across 23 orders.

Our analysis closely follows Sing et al. (2019)\citep[][]{sing:2019}, which we refer for further method details. 
We used the Jitter Detrending method to correct for time variable slit losses seen in the white light curve photometry, and fit the light curves using the system parameters given in Extended Data Table \ref{tab:params}. We find a band integrated white light curve $R_p/R_s$ of 0.1244$\pm$0.0050 which matches well (0.6-$\sigma$ significant difference) when compared to the same wavelength region as measured by UVIS ($R_p/R_s$=0.12133$\pm$ 0.00054). The main difference in our methods with ref \citep{sing:2019} was the use of a common-mode analysis when analysing the spectroscopic channels, where the best-fit transit model was removed from the white light curve raw photometry, and used to remove common instrument trends. In this case, large slit losses repeating every spacecraft orbit are seen in the photometry. The common-mode analysis removes the majority of the instrument trends seen, with the remaining modeled with a second order polynomial in HST orbital phase. The spectroscopic channels each reach a residual scatter that is consistent with the photon noise level. As done for WASP-121b, we divided the E230M spectra into 196 spectroscopic channels each with a 4\AA\ bandpass. The resulting spectrum can be seen in Extended Data Fig.~\ref{fig:E230M}. Compared to WASP-121b, WASP-178b does not show strong Fe II or Mg II absorption features, with the high resolution E230M spectrum consistent with the broadband NUV spectrum of UVIS.\\

\textbf{Atmosphere Models} Self-consistent 1-D PHOENIX atmosphere models of HAT-P-41b, WASP-121b, and WASP-178b were computed to compare the expected atmospheric temperatures in each planet to refractory species condensation curves. \replaced{The model setup was}{Model setups were} similar to past ultra-hot Jupiter studies with PHOENIX \citep{lothringer:2018b,lothringer:2019,lothringer:2020b}\replaced{ that}{, which} are computed on a 64-layer optical-depth grid from $\tau=$1e-10 to 1e2, which corresponds to similar magnitudes in pressure. \added{We ran models for two internal temperatures, 200 K and 700 K, which correspond to the lowest and highest internal temperatures expected for a hot Jupiter \citep{thorngren:2019}. We further assume full planet-wide heat redistribution to approximate temperatures at the terminator, consistent with similar investigations \citep{gao:2020}. The temperature structure from the self-consistent model of WASP-178b is quite similar to the retrieved temperature profile from PETRA (see Fig.~\ref{fig:pp_cond_comapre}), indicating the full heat redistribution assumption appears to be a good approximation of the\added{ average} conditions at the terminator.} 

The model includes opacity from 130 molecular species and neutral and ionized atomic species up to uranium. TiO and VO are important visible-wavelength opacity sources\citep{hubeny:2003,fortney:2008}, but observations suggest they do not always appear to be present in the atmosphere\citep{diamondlowe:2014}. Consistent with retrievals of observations from these planets \citep{lewis:2020,evans:2018}, the HAT-P-41b model did not include TiO or VO, the WASP-121b model only included VO, and the WASP-178b model only included TiO but at a reduced abundance (see Extended Data Table~\ref{tab:params} and Extended Data Fig.~\ref{fig:posterior}). All other abundances, including SiO and Fe, were treated in local chemical equilibrium. We further used the modelled HAT-P-41b transmission spectrum to rule out the presence of gaseous refractories on one or both limbs, further supporting the fact that these elements have rained out.
\\

\textbf{Atmosphere Retrievals} We used PETRA \citep[][]{lothringer:2020a} to retrieve atmospheric properties from the observations. PETRA uses a Differential-Evolution Markov Chain statistical framework \citep[][]{terbraak:2008} to sample the posterior distribution of the parameter space.\added{ Our retrieval setup was similar to previous transmission retrievals with PETRA \citep{lothringer:2021,wilson:2021}.} We parameterized the temperature structure using the 5-parameter approach of Parmentier \& Guillot (2014)\citep[][]{parmentier:2014}. We also included a necessary reference radius parameter. The abundances of major UV and optical opacity sources were treated as free parameters with uniform vertical abundance. These included Fe, Fe II, Mg, Mg II, TiO, VO, and SiO. We also included a free parameter for the metallicity ([Fe/H]) of the rest of the atmosphere, which was treated in chemical equilibrium. This allowed the effect of other potential absorbers expected to be \replaced{of lesser importance}{less important} (e.g., H$^-$, Ca, Ni, FeH) to be taken into account while reducing the number of free parameters to explore. Continuous opacity from H$^-$ and scattering from hydrogen and helium were also included. 

Uniform priors between volume mixing ratios of 10$^{-12}$ and 10$^{-1}$ were placed on each of the opacity sources. Priors were also placed on the temperature structure to avoid extremely low ($<500$~K) and extremely high ($>8,000$~K) temperatures. We ran a total of 120,000 iterations among 30 chains reaching a Gelman-Rubin $<$ 1.025. Extended Data Fig.~\ref{fig:posterior} shows 2-D cross-sections of the retrieved posterior distribution along with the 1-D marginalized distribution for each of the retrieved parameters. A summary of the retrieved atmospheric properties is included in Extended Data Table~\ref{tab:params}.

We also ran a series of retrievals without certain opacity sources in order to compare different atmospheric species' ability to fit the observed data. The scenarios we tested are listed in Table~\ref{tab:retrievals}. We computed the Bayesian Information Criteria for each scenario, taking the retrieval with all opacity sources included as our fiducial scenario (``Full") to calculate a $\Delta$BIC that quantifies whether there is statistical evidence to include a given parameter, in this case an opacity source, in the retrieval. Generally, a $\Delta$BIC between 2 and 6 indicates positive evidence for the inclusion of a given parameter, while $\Delta$BIC above 6 indicates strong evidence. Our retrieval analysis indicates there is strong evidence for the inclusion of SiO or Mg I, and thus the gaseous precursor species to silicate condensates. \added{We also ran a retrieval that included a haze parameterization \citep{macdonald:2017} to account for photochemical or other high-temperature aerosols, but found that they did not improve the fit.}

\added{In our fiducial scenario, the retrieval did find a highly super-solar abundance of Mg II. This result is being driven by the single data point at 0.28 microns (see Fig.~\ref{fig:modelcompare}). A retrieval without Mg II provides an similarly good fit to the data (see Table~\ref{tab:retrievals}), demonstrating that the inclusion of Mg II is not necessary to fit the data. We therefore do not choose to interpret the retrieved Mg II abundance as unambiguously physical. This is supported by the lack of any Mg II signal in the HST/STIS/E230M observations (see Extended Data Fig.~\ref{fig:E230M}).}
\\

\textbf{Stellar Activity} Starspots or faculae cause stellar inhomogeneities which can potentially contaminate transmission spectra \citep[][]{mccullough:2014,rackham:2018}. While magnetic activity (and thus starspots and facuale) are most relevant for low-mass stars, higher-mass stars may also show some degree of activity \citep[][]{rackham:2019} and some transmission spectra of hot Jupiters around early-type host stars may be consistent with unocculted stellar inhomogeneities \citep[][]{kirk:2021}. The problem is most acute at the shortest wavelengths where the flux\added{ contrast} between the nominal stellar photosphere and the active region is the greatest. 

We examined whether the transit spectrum of WASP-178b could be caused by unocculted stellar activity. We found that because of the magnitude of the spectral feature, extreme spot covering fractions and temperatures would be required. To fit the magnitude of the NUV feature, a starspot 1,500~K cooler than the nominal photosphere would require a spot covering fraction of 60\%. For a starspot 2,500~K cooler than the photosphere, a spot covering fraction of 50\% is required. Towards longer wavelengths, where the observed spectrum is flatter, the contaminated model spectrum continue sloping to small transit depths which is not seen in the data. In the end, these factors, combined with the satisfactory fit to the stellar SED without activity and low activity levels found in the star (\citep[see below and~][]{hellier:2019}), suggest that stellar activity cannot be responsible for the large feature in WASP-178b's transit spectrum.

The 2019 TESS light curve of WASP-178 shows a consistent 0.115\% variable photometric signal with a 0.185 day period\cite[][]{rodriguezmartinez:2020}, which is also easily visible in 2021 high cadence TESS photometry. This variability was speculated to be from $\delta$ Scuti pulations in WASP-178 and possible \replaced{gravity-darkening light curve asymmetries}{limb-asymmetries} were reported from the TESS data as well \cite[][]{rodriguezmartinez:2020}. However, our HST transit light curves show no evidence of any variability at the 0.115\% level, with the raw UVIS photometry showing variations less than 0.02\% over a 0.3 day window. Upon further inspection of the TESS field of view and near-by faint contaminant stars, from the ASAS-SN photometry database \citep{kockanek:2017,jayasinghe:2019}, we determine the origin of the photometric variations to be ASASSN-V J150908.07-424253.6 which is a nearby 14.5 magnitude W Ursae Majoris-type binary star with a reported period of 0.369526 days, which is an alias of the reported 0.185 day period.  Both the period and magnitude of variations match that of the signal seen as diluted in the TESS data by the brighter WASP-178b. As such, we conclude there is no evidence for WASP-178 to have any photometric variations larger than 0.02\%.  In addition, we find no evidence for reported transit asymmetries due to possible gravity-darkening effects\cite[][]{rodriguezmartinez:2020} in the HST data either. As a transit asymmetry signal in the TESS data could have also been influenced by the binary star, we analysed the 2021 TESS photometry taken at a higher cadence.  When removing the binary star contaminating variables, the TESS light curve shows no transit asymmetries in agreement with the HST data. Thus, there is no evidence that either gravity darkening nor significant photometric stellar activity are an issue with WASP-178.
\\

\textbf{Scattering} An alternative mechanism for producing large short-wavelength transit depths is through scattering. Small particles tend to scattering short-wavelength light more effectively than longer-wavelength light, leading to slopes towards greater transit depths at shorter wavelengths in transmission spectra \citep[][]{etangs:2008}. If we describe the scattering cross-section as:

\begin{equation}
    \sigma = \sigma_0(\lambda/\lambda_0)^{\alpha}
\end{equation}

the slope in the transmission spectrum can be expressed as:

\begin{equation}
    \frac{1}{H} \frac{\textrm{d}R_p }{\textrm{dln}(\lambda)} = \alpha
\end{equation}

Rayleigh scattering, the limit that the particle is smaller than the wavelength of light, has a characteristic $\alpha = -4$. Given our observed feature magnitude of approximately 20 scale heights over 0.2~\microns{}, we calculate $\alpha = -28.99$ is required to match the data. While super-Rayleigh slopes in transmission spectrum are possible due to a vertical gradient in opacity \citep[][]{ohno:2020}, a slope with $\alpha = -28.99$ would still be difficult to create, especially with non-purely scattering particles like a photochemical haze. \added{Additionally, we do not expect aerosols to survive to the pressures \added{or temperatures} that we probe with our observations ($\sim1$~microbar\added{, $\sim$~4000~K}).} 
\\

\textbf{\added{Limb Asymmetries}}
\added{Limb asymmetries can potentially complicate the interpretation of a transmission spectrum.  In particular, because of atmospheric advection from the hotter day side to cooler night side, the morning terminator can potentially be cooler with increased condensate `clouds' while the hotter evening terminator can be cloud-free. This effect may be evident in WASP-76b\cite{ehrenreich:2020}, and theoretical models have investigated the effect by coupling cloud formation to atmospheric dynamics\cite{powell:2019,helling:2021,parmentier:2021,roman:2021}. For WASP-178b, if the limb asymmetries were prevalent, we would expect the NUV transmission spectrum to be strongly affected, as features such as SiO could be in gaseous form on one limb, but condensed into aerosols on the other. The combined effect would be to potentially bias the interpretation toward the hotter-clearer limb albeit with a reduced signal.}  

\added{As our data covers some of both ingress and egress, where limb asymmetries have large observable effects, we searched the UVIS data between 0.18 and 0.28 $\mu$m for terminator asymmetries using \texttt{catwoman} \citep{espinoza:2021}. 
In a scenario where Mg and SiO are condensed on the leading morning terminator and has an effective radius consistent with the optical ($R_p/R_s$=0.11133$\pm$0.0005), the trailing evening terminator would require a radius of $R_p/R_s$=0.12924 in order to match the transit radius ratio of $R_p/R_s$=0.12062$\pm$0.00067 measured in the NUV with the +1 order. The NUV +1 order light curve and magnitude of limb asymmetries are shown in Extended Data Fig. \ref{fig:limb_asym}.  In ingress/egress, such asymmetries are detectable in the data as they are found to have a $\sim$500 ppm effect on the transit light curve, which is comparable to the 1-$\sigma$ error bars (450 ppm). This simplified model is ruled out at the 5-$\sigma$ level by the +1 order alone (see Extended Data Fig. \ref{fig:limb_asym}). To place further constraints, we used \texttt{catwoman} to fit the +1 and -1 order NUV data (0.18 to 0.28 $\mu$m) for the two hemisphere planetary radii, $R_{p,1}$ and $R_{p,2}$, as well as terminator inclination angle $\phi$. We found $\phi$ to be unconstrained and $R_{p,1}$ and $R_{p,2}$ were consistent at 1-$\sigma$, favoring a scenario without limb asymmetry. We also fixed $\phi$ to strictly assume an east/west limb asymmetry. In this case, we also find both hemispheres fit to nearly the same radii, $R_{p,1}/R_s$=0.1195$^{+0.0020}_{-0.0021}$ and $R_{p,2}/R_s$=0.1211$^{+0.0019}_{-0.0020}$, which are both larger than the optical radius at $>$3-$\sigma$ confidence.}

\added{This indicates \added{that }the NUV transmission spectral features (SiO/Mg) occur on both the leading and trailing limbs. With no indications either limb being cloudy, potential silicate condensates are confined to the night-side of the planet on WASP-178b. However, WFC3 phase-curve observations of the similar planet WASP-121b \citep{mikal-evans:2021} show the night-side temperatures do not generally drop low enough to be conducive of silicate material condensation as significant heat is transported.}

\added{We also note that the fact that both hemispheres are the same radius to within 1-$\sigma$ could point to implications for the atmospheric circulation and heat transport at the low pressures probed in transit at NUV wavelengths. If the atmospheric circulation was dominated by super-rotation at these pressures, one would expect the morning terminator to be much colder, and thus have a smaller radius at a given optical depth, than the evening terminator. Because this does not seem to be the case, our observations might be an indication that the circulation at microbar and less pressures is dominated by a day-to-night flow, whereby the morning and evening terminators would be more similar in temperature, and thus radius. This behavior is in line with theoretical expectations \citep{showman:2013b}.} 
\\


\textbf{Data Availability}
The raw data from this study is publicly available via the Space Science Telescope Insitute's Mikulski Archive for Space Telescopes (https://archive.stsci.edu/).
\\

\textbf{Code Availability}
The raw data was reduced with the available STScI CALWF3 pipeline and spectra were extracted with the public IRAF APALL routines. The light curve fitting used custom routines that we opt not to make public due to undocumented intricacies. Model and retrievals were generated using PHOENIX, which is a proprietary code but described in many publications \citep[e.g.,][]{hauschildt:1999, barman:2001}.
\\

\textbf{Acknowledgements}
We thank the reviewers for their helpful report which improved the paper. We thank the UV-SCOPE team for relevant discussion. We thank T. Barman for the use of a computing resources used in the calculation of the atmospheric retrievals. Support for this work was provided by NASA through grant number HST-GO-16086 from the Space Telescope Science Institute, which is operated by AURA, Inc., under NASA contract NAS 5-26555. This research has made use of the NASA Astrophysics Data System and the NASA Exoplanet Archive, which is operated by the California Institute of Technology, under contract with the National Aeronautics and Space Administration under the Exoplanet Exploration Program.
\\

\textbf{Author Contributions}
J.D.L. and D.K.S. contributed equally to this work. J.D.L led the observing proposal and retrieval analysis. D.K.S. led the data analysis with contributions from Z.R., H.R.W., J.J.S., and A.T.W. All authors discussed the data analysis and interpretation and commented on the manuscript.
\\

\textbf{Competing Interests} The authors declare no competing interests.

\clearpage

\renewcommand{\figurename}{\textbf{Extended Data Fig.}}
\setcounter{figure}{0}   

\renewcommand{\tablename}{\textbf{Extended Data Table}}
\setcounter{table}{0}  


\begin{figure}
    \centering
    \includegraphics[width = 0.95\linewidth]{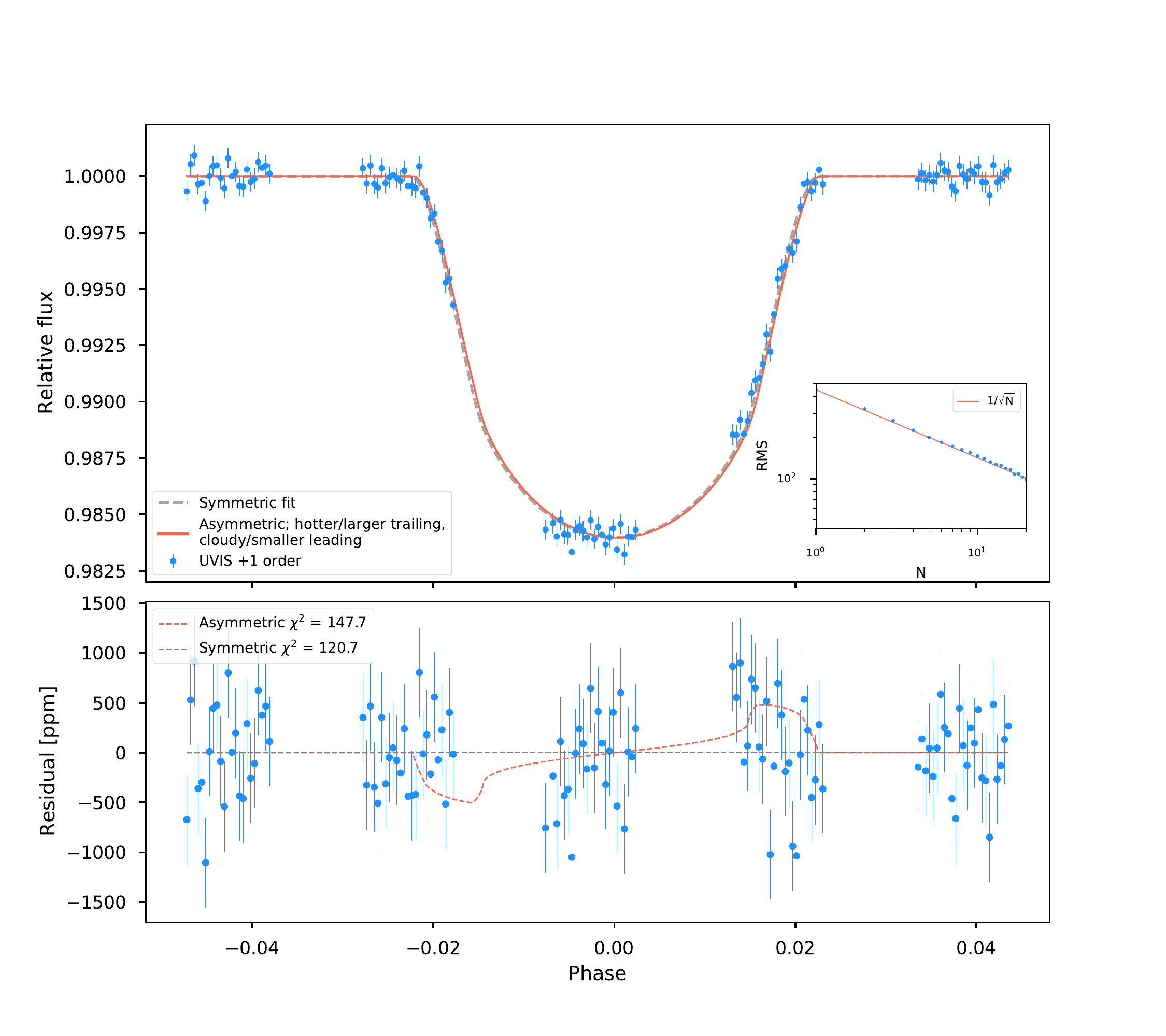}
    \caption{\textbf{WASP-178b Transit Asymmetry Analysis.} The 0.18-0.28 $\mu$m NUV light curve of WASP-178 b, with the best-fitting symmetric light curve, and an asymmetric light curve representing a scenario with a hotter/larger trailing terminator, and a colder/smaller leading terminator. The radius of the leading terminator was set to the optical value, and the trailing terminator was fixed to the value that fits the NUV transit depth. The inset shows the RMS scatter of the residuals as a function of number of points per bin, N. }
    \label{fig:limb_asym}
\end{figure}
\clearpage

\begin{figure*}[h]
	\centering
	\includegraphics[width=6.0in]{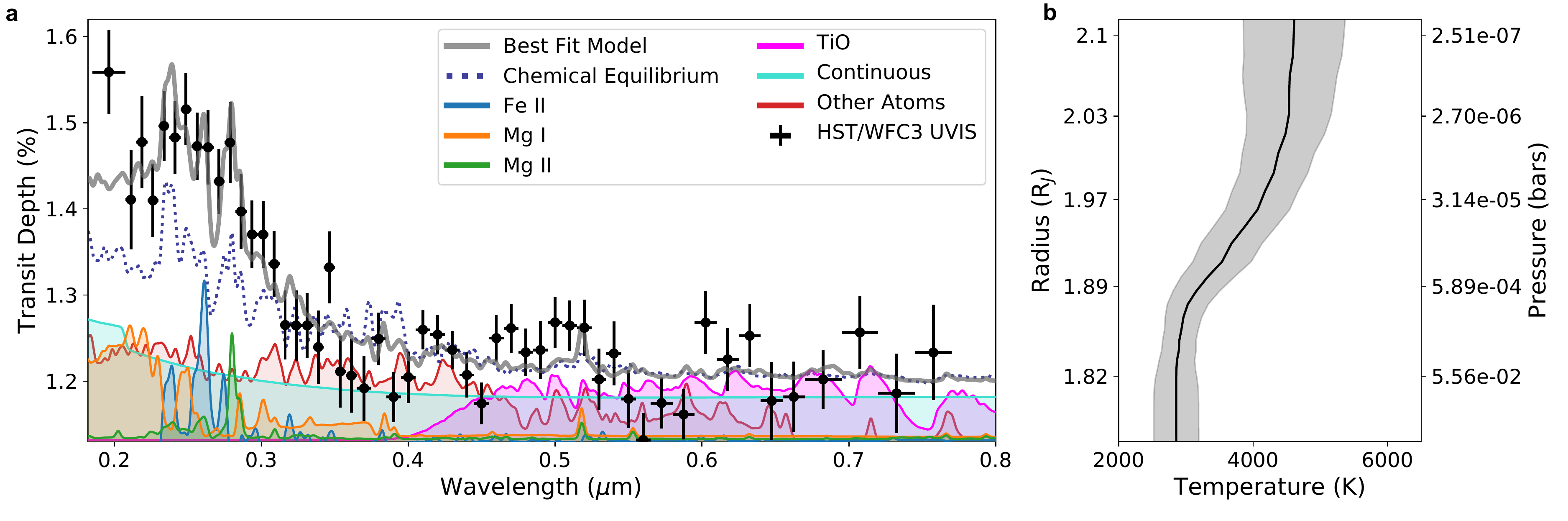} 
	\caption{\textbf{WASP-178b NUV-Optical Transmission Spectrum (No SiO)} Same as Fig.~\ref{fig:modelcompare}, but for the retrieval without SiO. Note the combined ability of Mg I and Fe II absorption to generate the large short-wavelength transit depths. \label{fig:modelcompare_nosio}}
\end{figure*}
\clearpage

\begin{figure*}[t]
	\centering
	\includegraphics[width=3in]{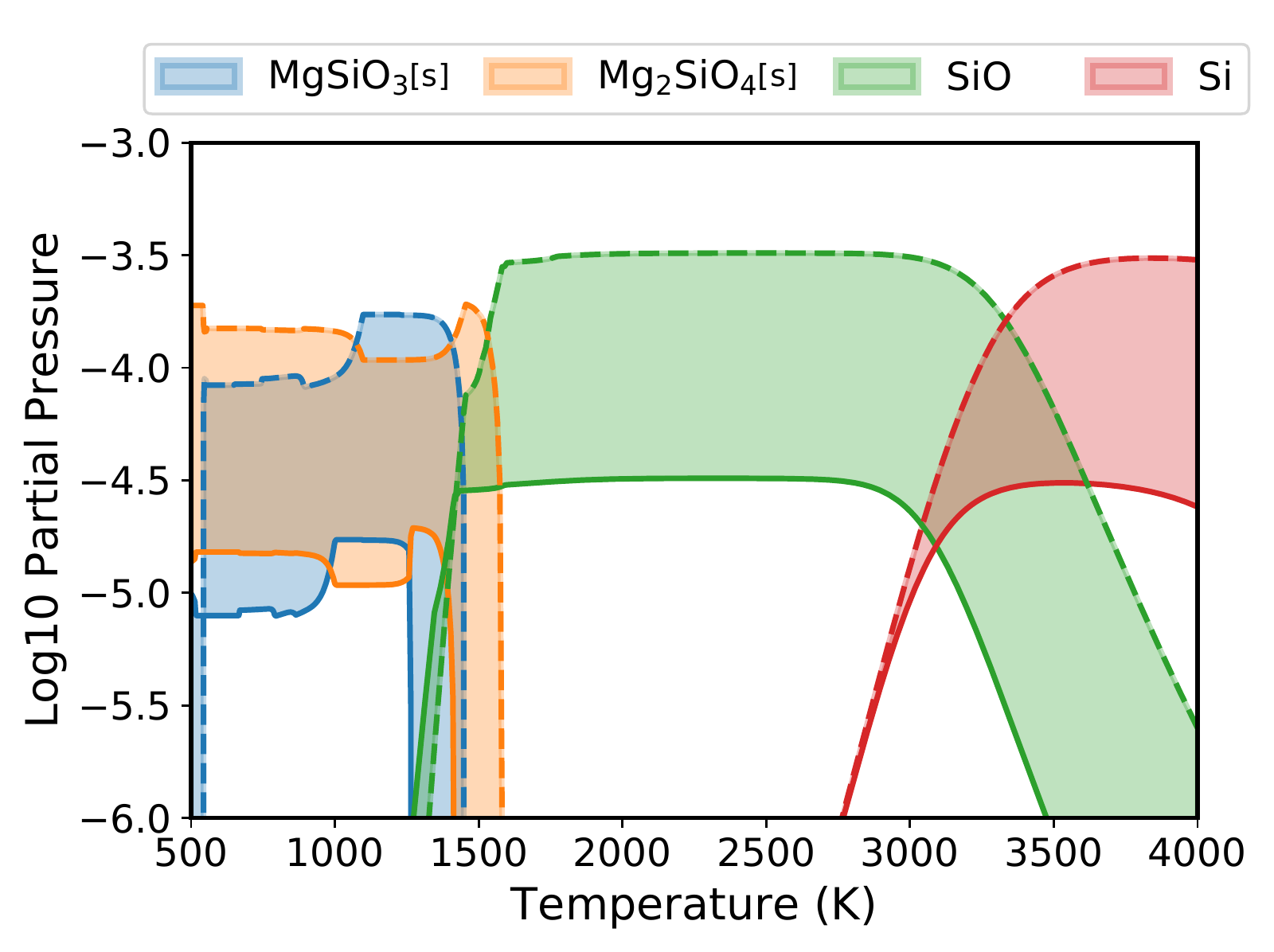} 
	\includegraphics[width=3in]{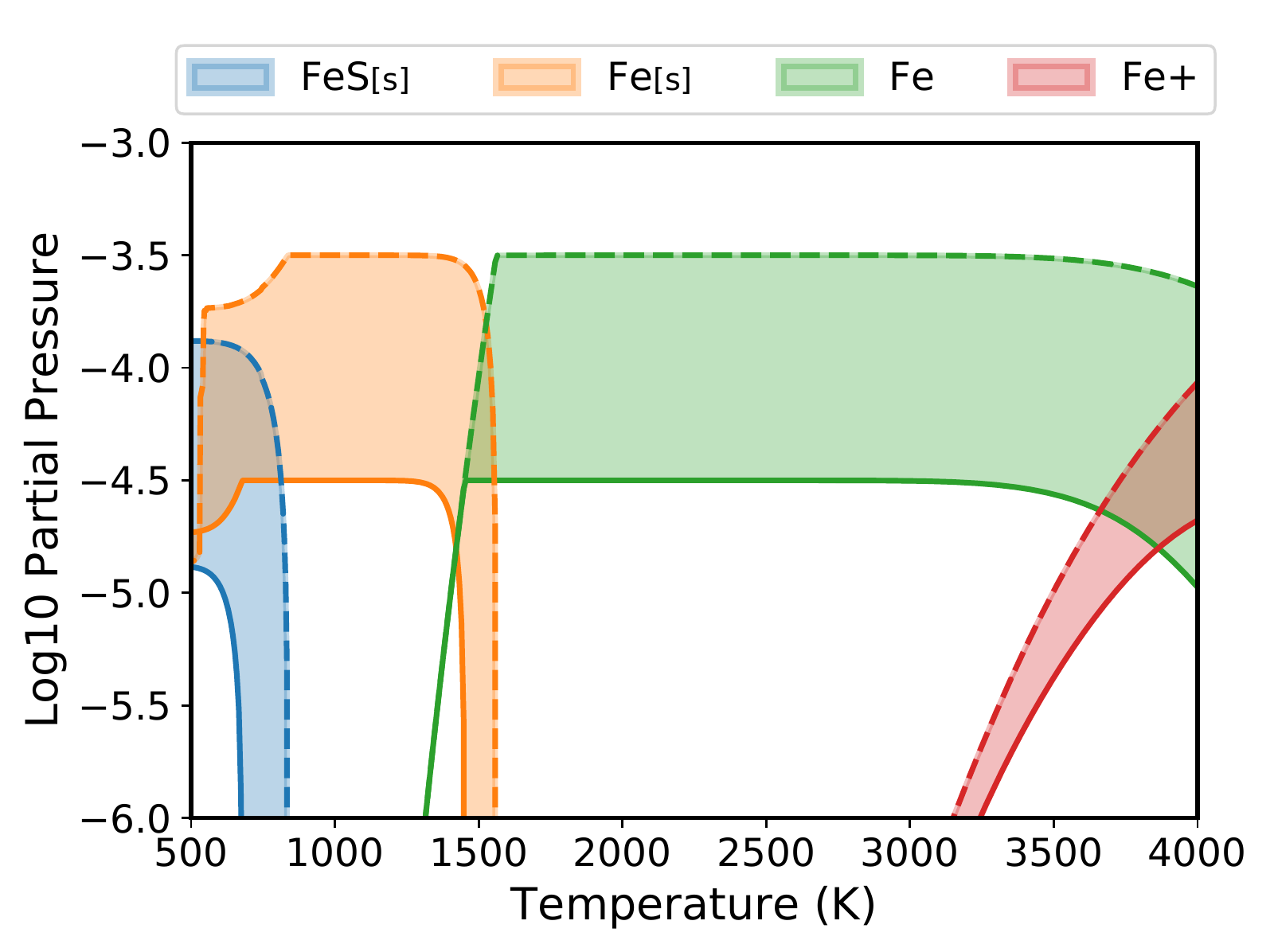} 
	\caption{\textbf{Chemical Equilibrium of Si and Fe.} Partial pressures of important silicon-bearing species (left) and iron-bearing species (right) at 1 mbar as a function of temperature. Equilibrium chemical abundances were calculated using GGchem \citep{woitke:2018}. \label{fig:si_pp}}
\end{figure*}
\clearpage

\begin{figure}[t]
	\centering
	\includegraphics[width=4.0in]{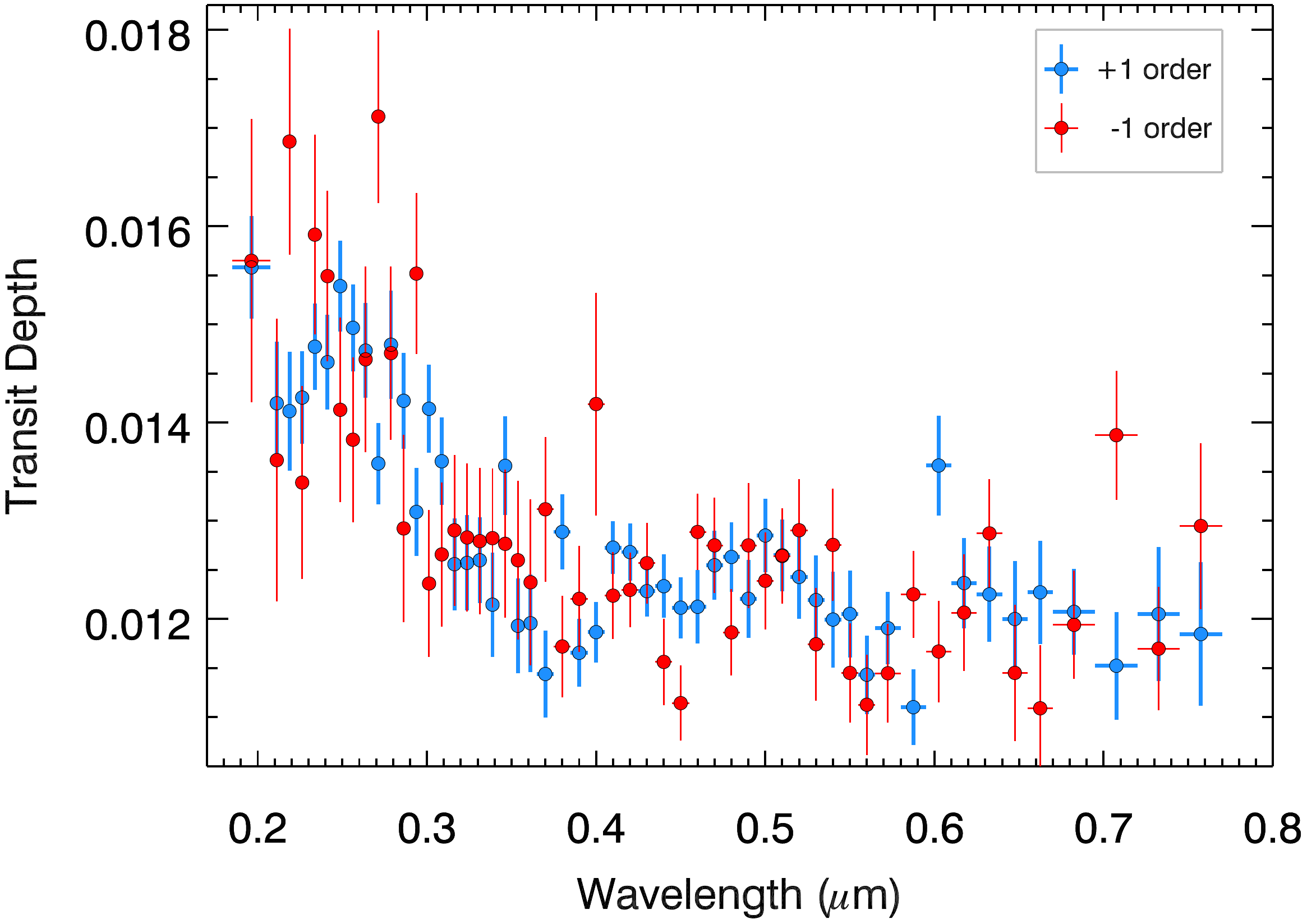} 
	\caption{\textbf{WASP-178b Spectral Order Comparison.} WFC3/UVIS G280 transmission spectrum of WASP-178b (with 1-$\sigma$ uncertainties) from the +1 (blue) and -1 order (red). The -1 order shows larger uncertainties due to a reduced throughput, but the transmission spectra show good agreement including an enhanced NUV absorption between 0.2 and 0.3 $\mu$m. \label{fig:orders}}
\end{figure}
\clearpage

\begin{table} 
\caption{WASP-178 HST/WFC3/UVIS transmission spectrum and noise properties.}
\label{Table:data}
\small
\begin{centering}
\begin{tabular}{cccccccccc}
\hline\hline  
&&\multicolumn{2}{c}{\underline{~~~~Combined~~~~}}&\multicolumn{3}{c}{\underline{~~~~~~~+1 order~~~~~~~}} & \multicolumn{3}{c}{\underline{~~~~~~~-1 order~~~~~~~}}\\
 $\lambda_c$ &$\Delta\lambda$ &  $(R_{P}/R_{*})^2$  &$\sigma_{(R_{P}/R_{*})^2}$ & $\sigma_w$ & $\sigma_r$ & $\beta$ & $\sigma_w$ & $\sigma_r$ & $\beta$\\
($\mu$m) & ($\mu$m) & (ppm) & (ppm)  & (ppm) & (ppm) & & (ppm) & (ppm) \\
\hline  
0.1963	&	0.0113	&	15587	&	490	& 1490  &	 0.	&	1.00	&	4344    &	0	&	1.00\\
0.2113	&	0.0037	&	14106	&	575	& 1809	&	287	&	1.06	&	4445	&	0	&	1.00\\
0.2188	&	0.0037	&	14775	&	538	& 1751	&	 0.	&	1.00	&	3355	&	0	&	1.00\\
0.2262	&	0.0037	&	14098	&	426	& 1376	&	 0.	&	1.00	&	2968	&	0	&	1.00\\
0.2338	&	0.0037	&	14962	&	405	& 1167	&	 0.	&	1.00	&	2937	&	0	&	1.00\\
0.2412	&	0.0037	&	14828	&	422	& 1333	&	 0.	&	1.00	&	2487	&	0	&	1.00\\
0.2488	&	0.0037	&	15155	&	417	& 1244	&	 0.	&	1.00	&	2697	&	0	&	1.00\\
0.2562	&	0.0037	&	14728	&	391	& 1202	&	 0.	&	1.00	&	2308	&	0	&	1.00\\
0.2637	&	0.0037	&	14714	&	430	& 1317	&	 0.	&	1.00	&	2692	&	0	&	1.00\\
0.2713	&	0.0037	&	14320	&	376	& 1116	&	 0.	&	1.00	&	2536	&	48	&	1.06\\
0.2788	&	0.0037	&	14768	&	469	& 1419	&	 0.	&	1.00	&	2555	&	0	&	1.00\\
0.2862	&	0.0037	&	13969	&	434	& 1314	&	 0.	&	1.00	&	2750	&	0	&	1.00\\
0.2937	&	0.0037	&	13703	&	394	& 1242	&	 0.	&	1.00	&	2275	&	0	&	1.00\\
0.3013	&	0.0037	&	13702	&	385	& 1215	&	 0.	&	1.00	&	2120	&	0	&	1.00\\
0.3088	&	0.0037	&	13360	&	381	& 1243	&	 0.	&	1.00	&	1943	&	0	&	1.00\\
0.3162	&	0.0037	&	12652	&	400	& 1263	&	 0.	&	1.00	&	2112	&	0	&	1.00\\
0.3237	&	0.0037	&	12646	&	408	& 1323	&	 0.	&	1.00	&	2089	&	0	&	1.00\\
0.3313	&	0.0037	&	12647	&	377	& 1169	&	 0.	&	1.00	&	2042	&	0	&	1.00\\
0.3388	&	0.0037	&	12396	&	425	& 1447	&	 0.	&	1.00	&	1944	&	0	&	1.00\\
0.3462	&	0.0037	&	13322	&	418	& 1308	&	 0.	&	1.00	&	2108	&	0	&	1.00\\
0.3537	&	0.0037	&	12112	&	416	& 1373	&	 0.	&	1.00	&	2247	&	0	&	1.00\\
0.3613	&	0.0037	&	12064	&	428	& 1361	&	 0.	&	1.00	&	2336	&	0	&	1.00\\
0.3700	&	0.0050	&	11918	&	379	& 1239	&	 0.	&	1.00	&	1991	&	0	&	1.00\\
0.3800	&	0.0050	&	12490	&	307	& 1155	&	 0.	&	1.00	&	1432	&	0	&	1.00\\
0.3900	&	0.0050	&	11817	&	291	&  985	&	 0.	&	1.00	&	1562	&	0	&	1.00\\
0.4000	&	0.0050	&	12047	&	299	&  930	&	 0.	&	1.00	&	2841	&	1142	&	1.31\\
0.4100	&	0.0050	&	12597	&	229	&  725	&	 0.	&	1.00	&	1255	&	0	&	1.00\\
0.4200	&	0.0050	&	12540	&	230	&  762	&	147	&	1.09	&	1103	&	0	&	1.00\\
0.4300	&	0.0050	&	12363	&	217	&  836	&	274	&	1.22	&	1163	&	0	&	1.00\\
0.4400	&	0.0050	&	12073	&	261	&  822	&	 0.	&	1.00	&	1268	&	0	&	1.00\\
0.4500	&	0.0050	&	11741	&	242	&  889	&	 0.	&	1.00	&	1110	&	0	&	1.00\\
0.4600	&	0.0050	&	12500	&	270	&  786	&	264	&	1.23	&	1115	&	0	&	1.00\\
0.4700	&	0.0050	&	12616	&	283	&  972	&	 0.	&	1.00	&	1309	&	220	&	1.07\\
0.4800	&	0.0050	&	12335	&	277	& 1076	&	172	&	1.06	&	1350	&	0	&	1.00\\
0.4900	&	0.0050	&	12361	&	338	& 1094	&	234	&	1.10	&	1833	&	0	&	1.00\\
0.5000	&	0.0050	&	12683	&	299	& 1010	&	 0.	&	1.00	&	1380	&	0	&	1.00\\
0.5100	&	0.0050	&	12645	&	292	&  956	&	 0.	&	1.00	&	1373	&	0	&	1.00\\
0.5200	&	0.0050	&	12620	&	329	& 1159	&	 0.	&	1.00	&	1382	&	0	&	1.00\\
0.5300	&	0.0050	&	12022	&	355	& 1219	&	 0.	&	1.00	&	1580	&	0	&	1.00\\
0.5400	&	0.0050	&	12322	&	371	& 1166	&	297	&	1.14	&	1472	&	291	&	1.09\\
0.5500	&	0.0050	&	11795	&	332	& 1187	&	 0.	&	1.00	&	1418	&	0	&	1.00\\
0.5600	&	0.0050	&	11316	&	316	& 1079	&	 49.	&	1.01	&	1464	&	0	&	1.00\\
0.5725	&	0.0075	&	11747	&	297	&  996	&	 0.	&	1.00	&	1440	&	0	&	1.00\\
0.5875	&	0.0075	&	11615	&	291	& 1020	&	132	&	1.04	&	1262	&	0	&	1.00\\
0.6025	&	0.0075	&	12680	&	363	& 1378	&	 0.	&	1.00	&	1471	&	0	&	1.00\\
0.6175	&	0.0075	&	12253	&	363	& 1221	&	165	&	1.04	&	1552	&	0	&	1.00\\
0.6325	&	0.0075	&	12527	&	364	& 1306	&	 0.	&	1.00	&	1512	&	0	&	1.00\\
0.6475	&	0.0075	&	11772	&	451	& 1575	&	 0.	&	1.00	&	1941	&	63	&	1.00\\
0.6625	&	0.0075	&	11818	&	407	& 1383	&	 0.	&	1.00	&	1778	&	0	&	1.00\\
0.6825	&	0.0125	&	12021	&	343	& 1136	&	 0.	&	1.00	&	1439	&	0	&	1.00\\
0.7075	&	0.0125	&	12566	&	423	& 1425	&	 0.	&	1.00	&	1739	&	0	&	1.00\\
0.7325	&	0.0125	&	11859	&	462	& 1735  &   336	&	1.09	&	1724	&	0	&	1.00\\
0.7575	&	0.0125	&	12333	&	553	& 1927	&	 0.	&	1.00	&	2299	&	260	&	1.03\\
\hline
\end{tabular}
\end{centering}
\end{table}
\clearpage

\begin{figure*}[th]
	\centering
	\includegraphics[width=6.0in]{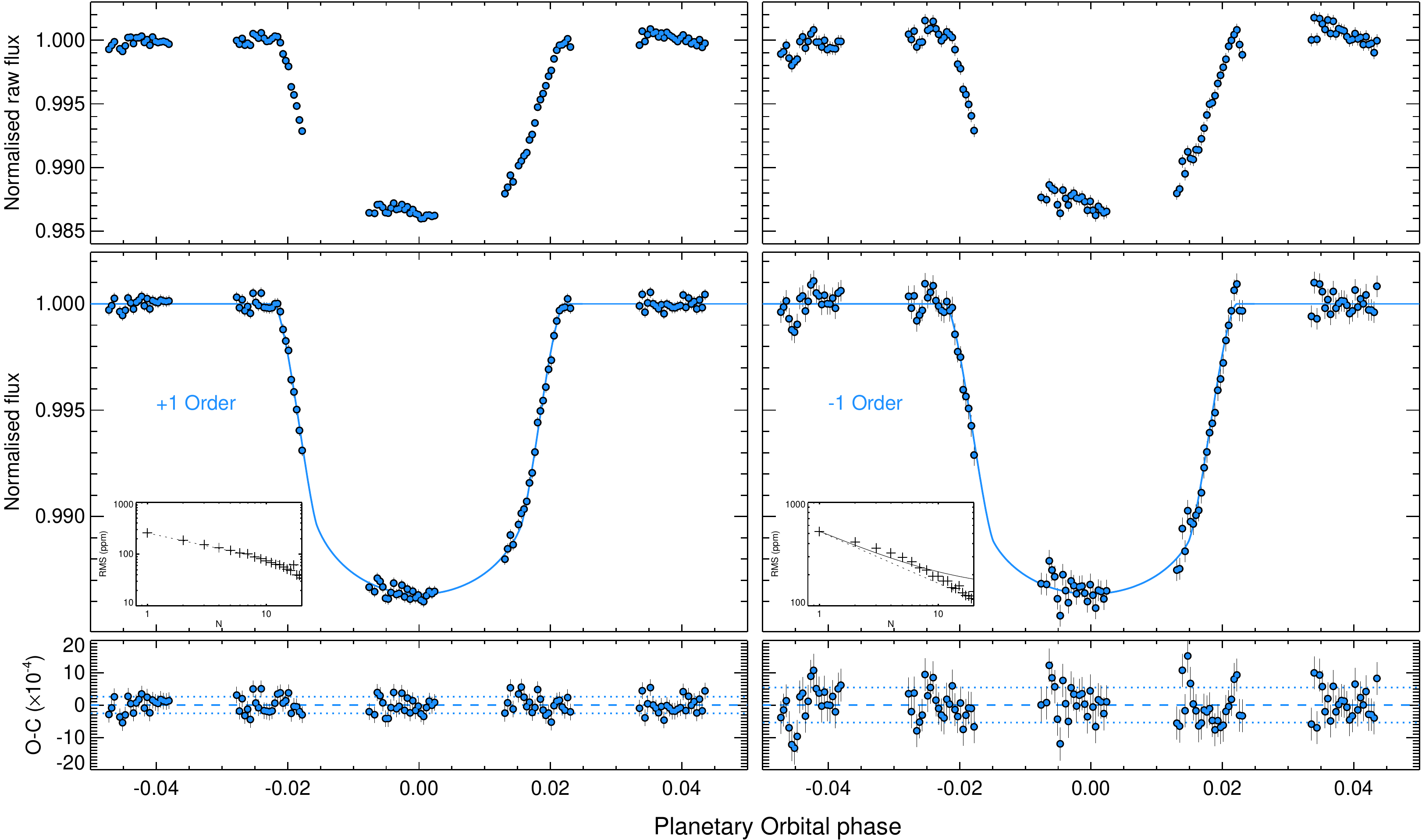} 
	\caption{\textbf{White light curves of the WASP-178b HST/WFC3-UVIS/G280 transit.} Error bars show the 1-$\sigma$ uncertainties. The left column shows the +1 spectral order, while the right column shows the -1 spectral order. The top row are the raw light curves the middle row are the light curves with systematics removed and a transit fit, and the bottom row are the residuals with the standard deviation of the residuals also shown (dotted lines). \added{Plots of the binned residual RMS are also shown.} \label{fig:wlc}}
\end{figure*}
\clearpage

\begin{figure*}[t]
	\centering
	\includegraphics[width=6.0in]{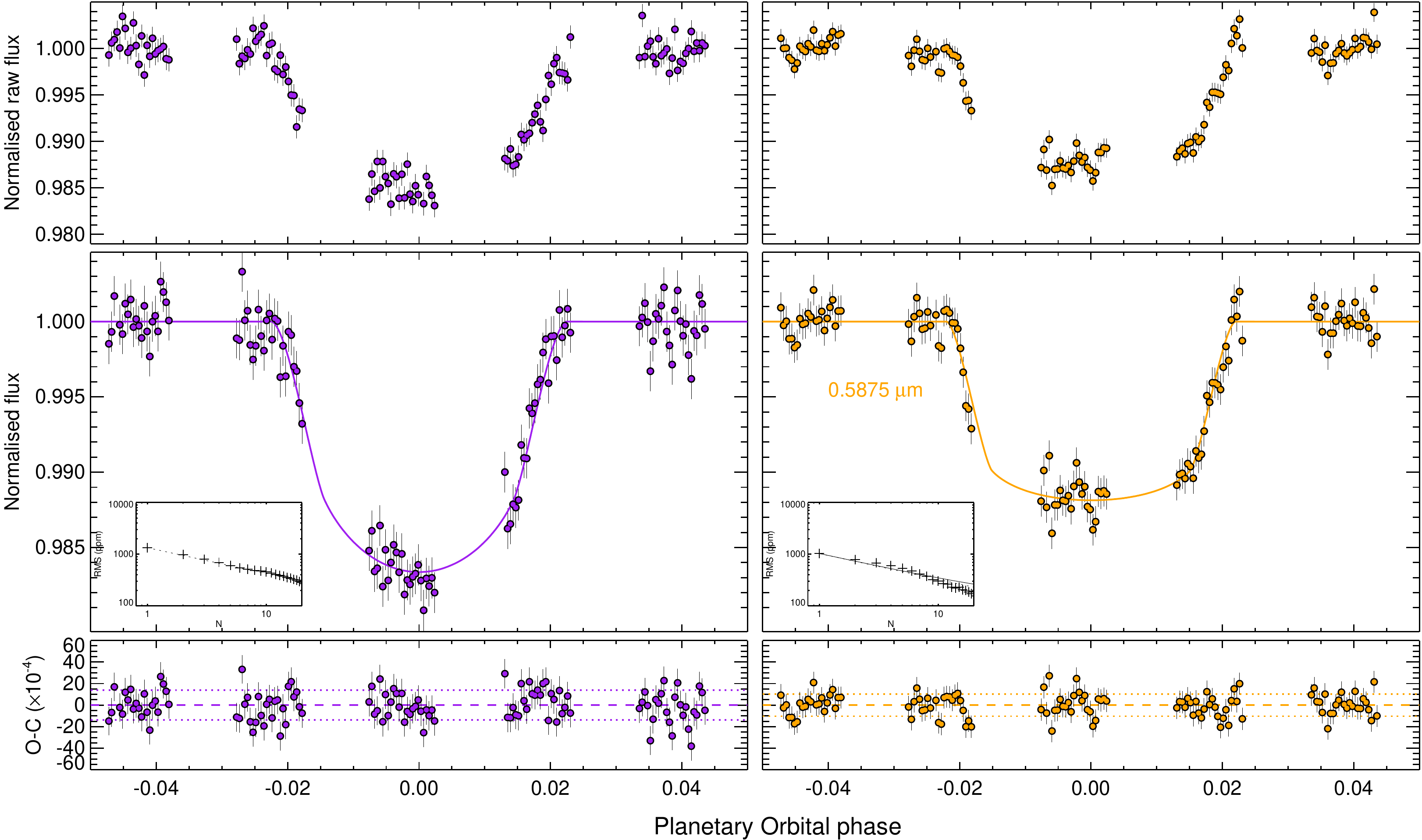} 
	\caption{\textbf{WASP-178b NUV-Optical Light Curve Comparison.} Two example fitted light curves from the +1 spectral order order from spectroscopic bins covering 0.2412 and 0.5875, with transit depths of 1.48 $\pm 0.04$\% and 1.16 $\pm 0.03$\%, respectively. The rows are the same as in Extended Data Fig.~\ref{fig:wlc}.  \label{fig:lcbins}}
\end{figure*}
\clearpage

\begin{figure*}[t]
	\centering
	\includegraphics[width=3.27in]{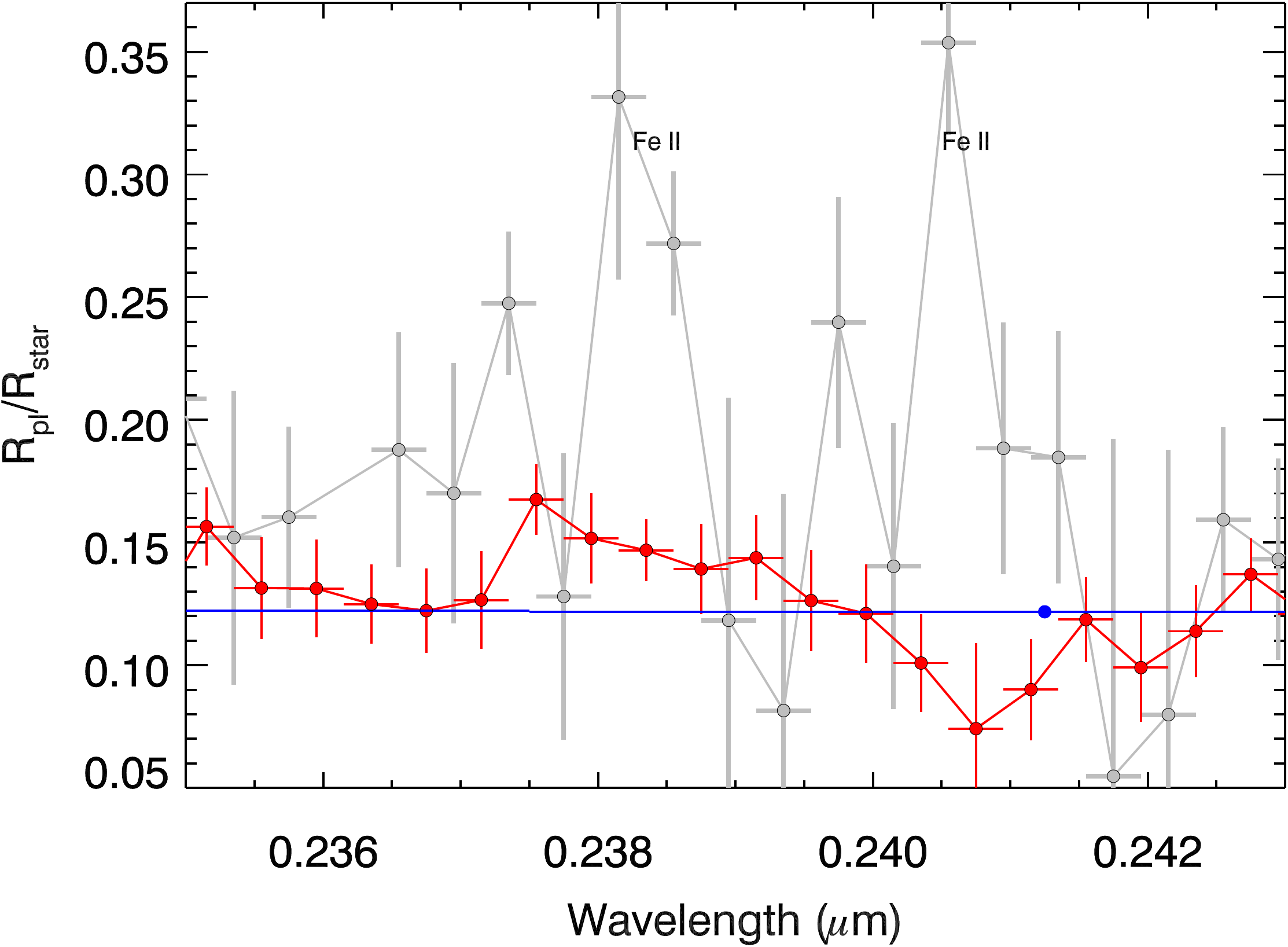} 
	\includegraphics[width=2.8in]{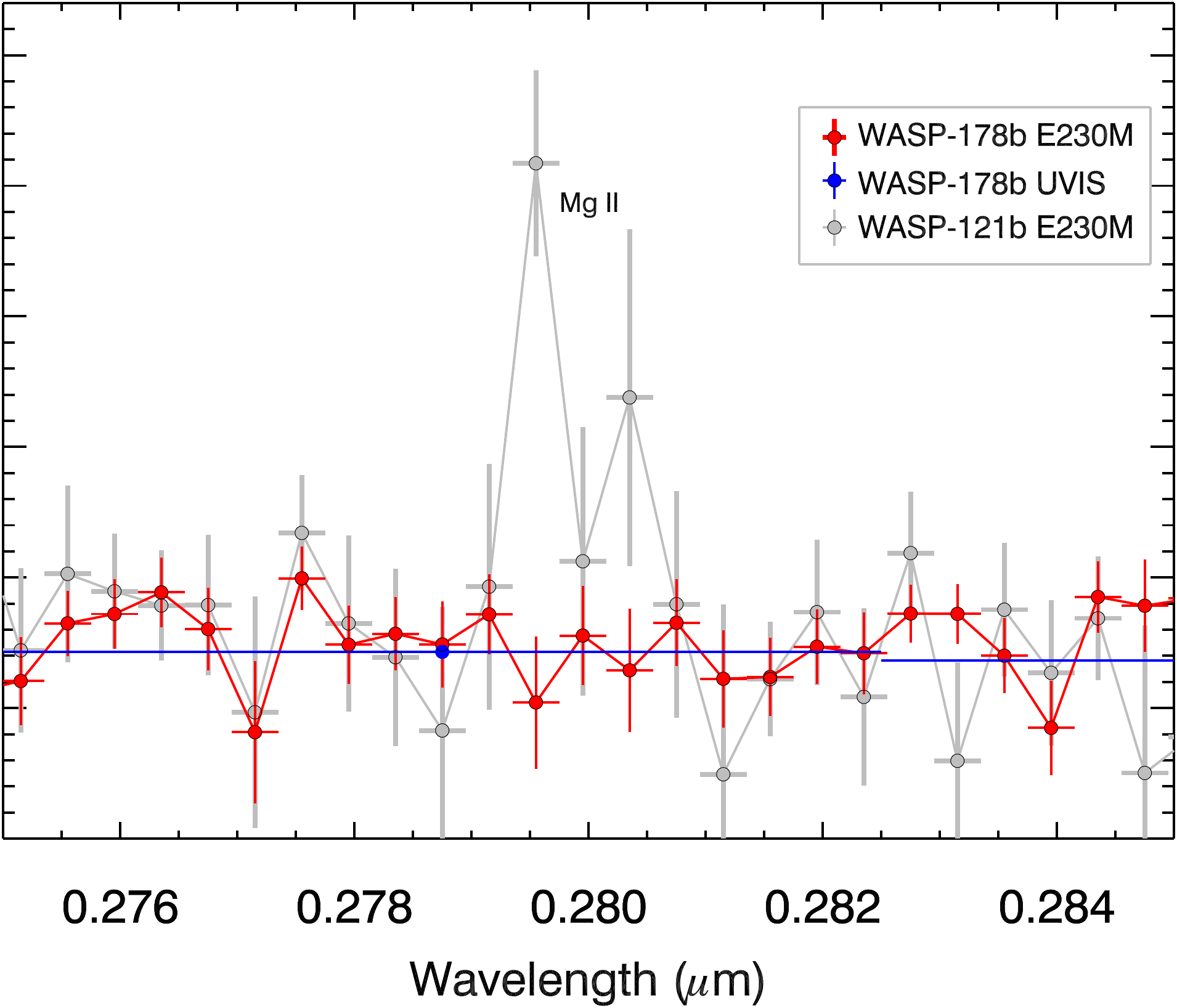} 
	\caption{\textbf{High-resolution HST/STIS/E230M Transmission Spectrum of WASP-178b.} NUV high resolution transit spectra of WASP-178b (with 1-$\sigma$ uncertainties) compared to WASP-121b around the Fe II (left) and Mg II lines. Shown are the spectra from STIS E230M for WASP-178b (red), WASP-121b\cite{sing:2019} (grey), and the low resolution UVIS spectra (blue). While WASP-121b shows strong Fe II and Mg II absorption features, the WASP-178b E230M spectra is consistent with the broadband NUV continuum with no Fe II or Mg II. \label{fig:E230M}}
\end{figure*}
\clearpage

	\begin{table*}[t] 
		\centering  
\caption{WASP-178b Fitted and Retrieved Orbital and Atmosphere Parameters \newline \footnotesize{*Equilibrium mixing ratios are shown in brackets and are calculated at WASP-178b's equilibrium temperature (2,450~K) and 0.1 mbar pressure.}}
		\label{tab:params} 
		\begin{tabular}{cc}
			\hline    
			\hline 
\multicolumn{2}{>{\centering\arraybackslash\hspace{0pt}}m{0.8\linewidth}}{\textbf{Orbital Parameters}} \\ 
\hline

Inclination ($i$) & $84.41\pm0.20^\circ$ \\ 
\hline
$a/R_{s}$ & $6.588\pm0.091$ \\ 
\hline
Transit Center ($T_0$) & $2459097.869279\pm0.00014$ days \\ 
\hline
Period \cite{hellier:2019} & 3.3448285$\pm$0.0000012 days\\
\hline
Planet Radius ($R_p/R_s$) & $0.11295150\pm0.00041$ \\ 
\hline
\hline

\multicolumn{2}{>{\centering\arraybackslash\hspace{0pt}}m{0.8\linewidth}}{\textbf{Retrieved and Equilibrium Atmospheric Log Mixing Ratios}} \\ 
\hline

SiO & $-4.68\pm^{0.65}_{0.59}$ [-4.46]\\ 
\hline
Fe  & $-9.51\pm^{1.64}_{1.49}$ [-4.47]\\ 
\hline
Fe II & $-3.76\pm^{1.35}_{4.15}$ [-9.44]\\ 
\hline
Mg & $-7.44\pm^{2.67}_{2.73}$ [-4.37]\\ 
\hline
Mg II & $-2.22\pm^{0.33}_{0.52}$ [-8.67]\\ 
\hline
TiO & $-10.01\pm^{0.59}_{0.72}$ [-7.07]\\ 
\hline
VO & $-10.76\pm^{1.01}_{0.79}$ [-8.56]\\ 
\hline
{[}Fe/H] & $-1.39\pm^{0.50}_{0.37}$ [0.00]\\
\hline
		\end{tabular}

	\end{table*} 
\clearpage

\begin{figure*}[t]
	\centering
	\includegraphics[width=6.5in]{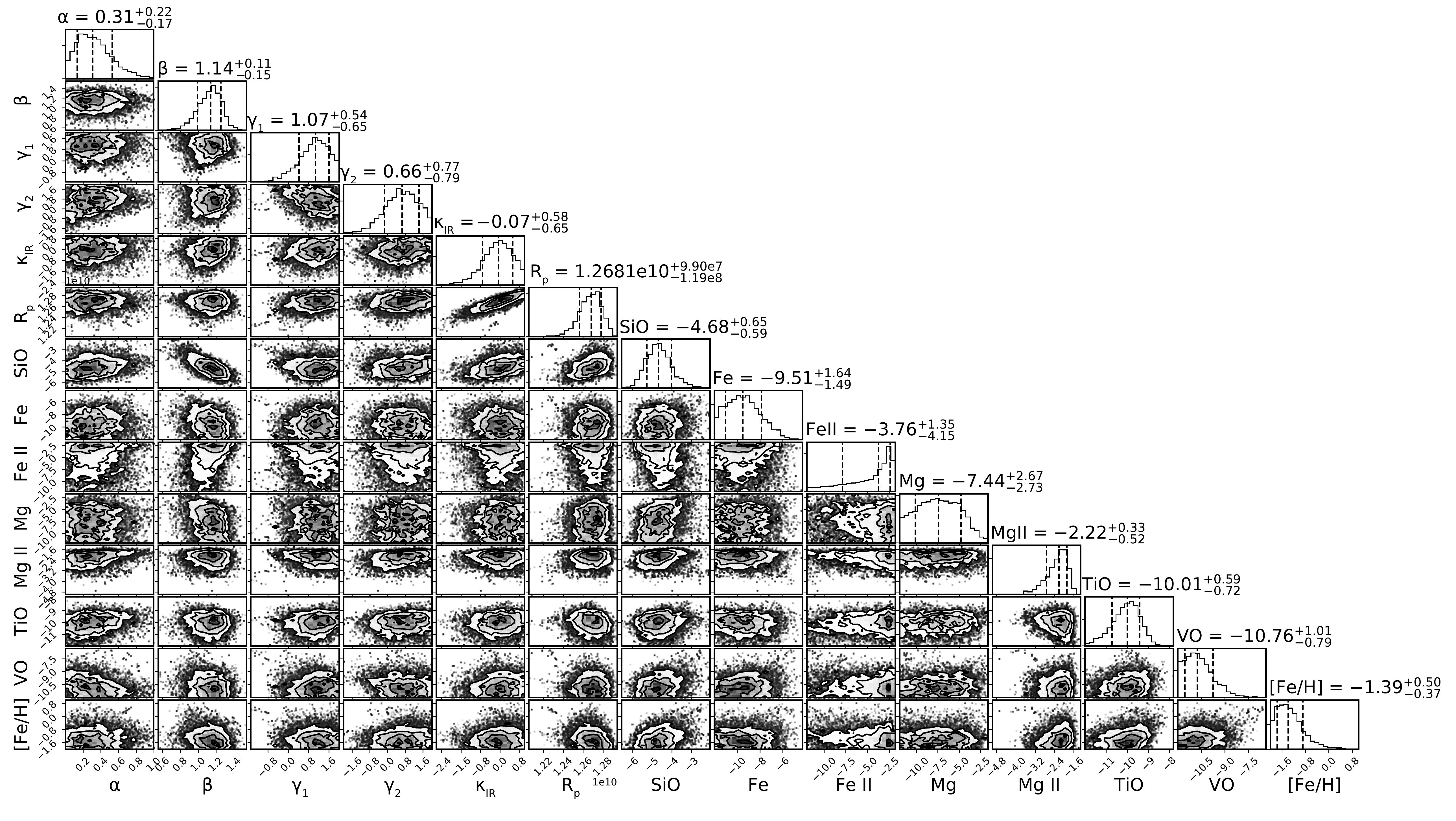} 
	\caption{\textbf{WASP-178b Atmospheric Retrieval Posterior Distribution.} 2-D cross-sections of the retrieved posterior distribution with 1-D marginalized distribution for the fitted parameters. The quoted quantities are the mean and 1-$\sigma$ retrieved values. The first five parameters are the temperature structure parameterization from Parmentier \& Guillot 2014 \citep{parmentier:2014}, the sixth is the reference radius, and the final eight are the various atomic and molecular abundances. \label{fig:posterior}}
\end{figure*}
\clearpage

\bibliographystyle{naturemag_noeprint}

\end{document}